\documentclass[a4paper,11pt,aps,prd,amsfonts,amssymb,amsmath,reprint,nofootinbib,twoside,floatfix]{revtex4-1}
\usepackage[english]{babel}
\usepackage[utf8]{inputenc}
\usepackage[colorinlistoftodos, color=green!40, prependcaption]{todonotes}

\usepackage{amsthm}
\usepackage{mathtools}
\usepackage{physics}
\usepackage{xcolor}
\usepackage{graphicx}
\usepackage{adjustbox}
\usepackage{placeins}
\usepackage[T1]{fontenc}
\usepackage{lipsum}
\usepackage{csquotes}
\usepackage{fancyhdr}
\usepackage[
left=20mm,
right=20mm,
top=25mm,
bottom=25mm,
columnsep=20pt
]{geometry}
\usepackage[pdftex, pdftitle={Article}, pdfauthor={Author}]{hyperref} % For hyperlinks in the PDF
\begin{document}
\title{Observational Constraints on WIMP Mini-Spikes Around Stellar-Mass Primordial Black Holes with 17 Years of Fermi-LAT Data}

\author{Ana Vitoria de Almeida Martinheira Braga }
\email{ana.martinheira@ifsc.usp.br }
\affiliation{Instituto de Física de São Carlos, Universidade de São Paulo, IFSC – USP, 13566-590, São Carlos, SP, Brasil}

\author{Murillo Gregorio Grefener da Silva }
\email{murilloggsilva@usp.br }
\affiliation{Instituto de Física de São Carlos, Universidade de São Paulo, IFSC – USP, 13566-590, São Carlos, SP, Brasil}

\author{Aion Viana}
\email{aion.viana@ifsc.usp.br}
\affiliation{Instituto de Física de São Carlos, Universidade de São Paulo, IFSC – USP, 13566-590, São Carlos, SP, Brasil}
\date{\today} % Leave empty to omit a date

\begin{abstract}
The quest to identify the true nature of dark matter remains one of the most pressing challenges in modern physics. We present a novel approach to probe the Weakly Interacting Massive Particle (WIMP) paradigm by analyzing density enhancements, or ``mini-spikes,'' around stellar-mass black holes (sBHs) using 17 years of data from the \textit{Fermi} Large Area Telescope. Motivated by the anomalous orbital decay observed in the black hole low-mass X-ray binaries A0620--00 and XTE J1118+480, we model these systems under the hypothesis of adiabatic spike formation around primordial black holes, incorporating the effects of tidal disruption in the Galactic disk. Finding no statistically significant gamma-ray excess at either location ($TS < 1$), we derive 95\% C.L. upper limits on the WIMP annihilation cross section. Our results exclude the canonical thermal relic cross section ($3 \times 10^{-26} \, \text{cm}^3\text{s}^{-1}$) across the 10~GeV to 10~TeV mass range for $b\bar{b}$ and $W^+W^-$ channels, and up to $\sim$6~TeV for the $\tau^+\tau^-$ channel. Recasting these results into a Galactic discovery reach, we demonstrate that \textit{Fermi}-LAT is sensitive to $10\,M_\odot$ mini-spikes even at distances surpassing the Galactic Center, provided the WIMP mass is below 1~TeV. These findings establish a significant tension between the dynamical friction interpretation of orbital decay in these systems and the WIMP hypothesis, providing robust observational constraints on the coexistence of primordial black holes and annihilating dark matter.
\end{abstract}

\maketitle

\section{Introduction}

The true nature of dark matter (DM) remains one of the fundamental and unresolved problems in modern physics. Among the most motivated candidates is the Weakly Interacting Massive Particle (WIMP), a thermal relic from the early Universe with a weak-scale interaction cross section~\cite{Arcadi:2017kky,Arcadi:2024ukq}. In the standard paradigm, WIMPs can self-annihilate into Standard Model particles, such as gamma rays, neutrinos, and cosmic rays. Indirect detection efforts prioritize regions of high DM density, where the annihilation rate, scaling with the square of the density ($\rho^2$), is significantly amplified~\cite{Slatyer:2021qgc}.

While dwarf spheroidal galaxies and the Galactic Center are traditional targets for such searches, black holes (BHs) provide a unique laboratory for probing the WIMP hypothesis. The adiabatic growth of a BH within a DM halo induces a process known as adiabatic contraction, reshaping the local distribution into a steep, ultracompact \textit{mini-spike}~\cite{gondolo1999dark,Sadeghian:2013laa}. These structures can boost the gamma-ray luminosity by many orders of magnitude, potentially making even stellar-mass black holes (sBHs) detectable as point-like gamma-ray sources.

Recently, observations of the black hole low-mass X-ray binaries (BH-LMXBs) A0620--00 and XTE J1118+480 revealed anomalous orbital decay rates that exceed predictions from magnetic braking, mass transfer from the companion star to the BH, and standard gravitational-wave emission~\cite{2013AJ....145...21K,vanGrunsven:2017nua,Neilsen:2007pb,Hernandez:2013haa,2017MNRAS.465L..15G}. This discrepancy has been interpreted as evidence for dynamical friction within dense DM spikes~\cite{chan2023indirect}. If these sBHs are primordial in origin, the formation of such spikes is a robust prediction of their evolution in a WIMP-rich environment~\cite{1985ApJS...58...39B, Mack:2006gz, Ricotti:2007jk, ireland2025dark}.

%Seminal studies indicate that black holes embedded within dark matter halos significantly perturb the local distribution (e.g., Gondolo \& Silk 1999; Gnedin \& Primack 2004; Fields et al. 2014; Shapiro \& Shelton 2016). This mechanism, governed by the conservation of adiabatic invariants, drives the assembly of ultracompact cusps. Under the WIMP hypothesis, stellar-mass remnants of a primordial nature are expected to accrete these steep substructures. Given that the annihilation signature scales as the profile ($\rho^2$), the resulting luminosity is predicted to be dramatically amplified.

In this paper, we analyze 17 years of \textit{Fermi} Large Area Telescope (LAT) data to search for gamma-ray signatures from these systems. By employing the most recent background models and source catalogs to perform a standard statistical analysis, we found no significant excess at the position of the sBHs. Under the assumption of WIMP annihilation flux within the proposed spikes, even accounting for historical tidal disruption, we derive extremely stringent constraints on the WIMP-sBH paradigm and the primordial origin of these systems.

%Consequently, we provide constraints on Dark Matter (DM) properties that narrow the allowed parameter space for proposed models. These limits exclude the canonical thermal-relic cross section for DM masses ranging from the GeV scale to the multi-TeV regime, depending on the annihilation channel. We further recast the Fermi-LAT flux sensitivity into a lower-limit distance reach for hypothetical stellar-mass black holes across representative sky directions, providing a quantitative estimate of when WIMP mini-spikes should become detectable by current gamma-ray observations. Finally, we discuss the implications for a mixed DM model consisting of WIMPs and Primordial Black Holes (PBHs).
  %I believe leaving the sections in separate files is more organized, change it if you desire 
\section{Dark Matter Spikes Around Stellar-Mass Black Holes}

The adiabatic growth of a BH within a DM halo reshapes the local density into a steep, ultracompact spike~\cite{gondolo1999dark}. Provided the gravitational potential evolves slowly relative to DM orbital timescales, an initial density profile $\rho(r) \propto r^{-\gamma}$ transforms into a spike $\rho(r) \propto r^{-\gamma_\mathrm{sp}}$, where the spike index is given by $\gamma_\mathrm{sp} = (9-2\gamma)/(4-\gamma)$~\cite{gondolo1999dark}. Following the framework in Ref.~\cite{ireland2025dark}, if these BHs are primordial in origin, they are expected to accrete DM from a nearly uniform background during their formation and subsequent evolution. Consequently, we adopt an initial slope $\gamma \approx 0$, which leads to a spike index $\gamma_\mathrm{sp} = 2.25$ under the adiabatic approximation~\cite{gondolo1999dark, ireland2025dark}. 

We assume these mini-spikes are currently embedded in a local Galactic DM distribution described by an Einasto profile,
\begin{equation}
  \rho_{\mathrm{Ein}}(r)=\rho_s\exp\!\left[-\frac{2}{\alpha}\left(\left(\frac{r}{r_s}\right)^\alpha-1\right)\right],
\end{equation}
with parameters $\alpha=0.17$, $r_s=20$~kpc, and $\rho_s=0.081$~$\mathrm{GeV/cm^{3}}$~\cite{Pieri:2009je, CTA:2020qlo}. The profile is normalized to yield $\rho_\odot=0.39$~$\mathrm{GeV/cm^{3}}$ at the Galactocentric radius $R_\odot=8.5$~kpc. %\MG{(Aion, na minha cabeça seria necessária uma REF com esses valores, mas eu nunca a encontrei. Os parâmetros do perfil Einasto que usei por todo esse tempo foram os que vc enviou para mim e para o Felipe no primeiro email sobre esse assunto de sBHs, mas nunca entendi sua escolha. Se usarmos os valores de parametrização do Cirelli (iopscience.iop.org/article/10.1088/1475-7516/2011/03/051/), por exemplo, obtemos uma densidade local de 0.3 a 8.33 kpc e 0.29 a 8.5 kpc.)}

\subsection{Density Profile and Spike Survivability}

The dark matter density profile surrounding the sBHs is governed by the gravitational influence of the BH, the limiting effects of WIMP self-annihilation, and relativistic truncation. We parametrize the spike structure as
\begin{equation}
    \rho(r) = \begin{cases}
        \rho_\mathrm{loc}, &  r \ge R_\mathrm{sp} \\
        \rho_\mathrm{loc} (R_\mathrm{sp}/r)^{\gamma_\mathrm{sp}}, & R_\mathrm{sp} \ge r \ge R_\mathrm{sat} \\
        \rho_\mathrm{sat} (R_\mathrm{sat}/r)^{0.5}, & R_\mathrm{sat} \ge r \ge 2R_\mathrm{Sch} \\
        0, & r \le 2R_\mathrm{Sch}
    \end{cases},
\end{equation}
where $R_\mathrm{sp}$ is the spike radius, defined as the region where the DM mass equals twice the BH mass~\cite{Merritt:2003qc}, and $R_\mathrm{Sch}$ is the Schwarzschild radius.

\begin{figure}[htbp]
    \centering
    \includegraphics[width=0.5\textwidth]{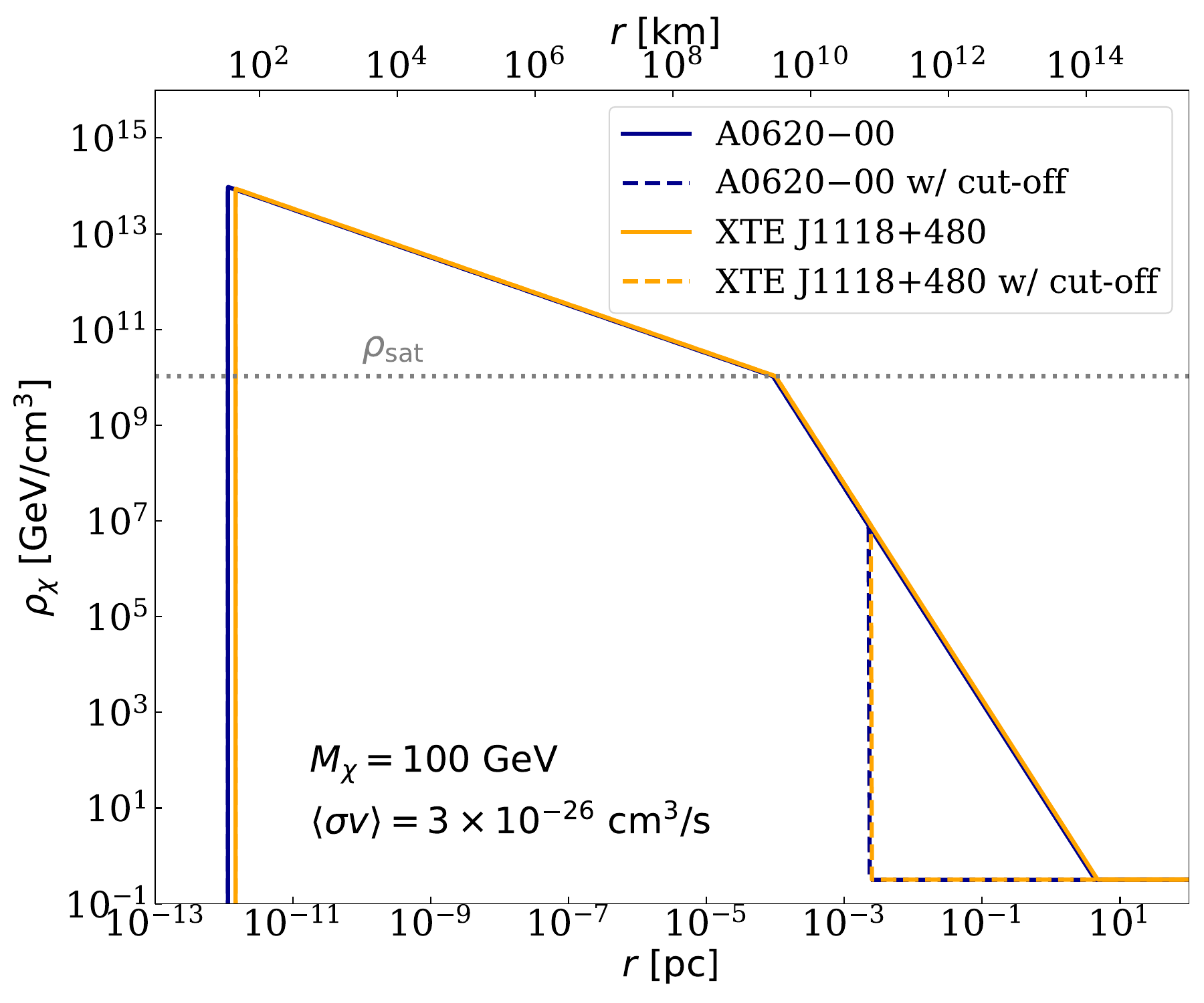}
    \caption{Dark matter density profiles around the stellar-mass black holes A0620--00 and XTE~J1118+480 under the spike formation hypothesis, shown both with and without the effects of tidal disruption on the subhalo. The gray dotted line marks the saturation density corresponding to the adopted values of DM particle mass and annihilation cross section shown in the figure, and a system lifetime of $10^{10}$~yr.}
    \label{fig:rho_StMBHs}
    % Gerado em:
    % Python_Murillo/micro_dm_spikes/j_factor(m&sigmav).ipynb
\end{figure}

The growth of the spike is regulated by the saturation density, $\rho_\mathrm{sat}$, above which DM self-annihilation depletes the density over the system lifetime $\tau_\mathrm{BH}$. Following Ref.~\cite{sadeghian2013dark}, to account for the non-circular nature of DM trajectories near the horizon, the distribution inside the saturation radius $R_\mathrm{sat}$ flattens into a weak cusp with slope $\gamma_\mathrm{sat} = 0.5$. These quantities are defined as
\begin{equation}
    \rho_\mathrm{sat} = \frac{M_\chi}{\langle \sigma v \rangle \tau_\mathrm{BH}}, \quad R_\mathrm{sat} = R_\mathrm{sp} \left( \frac{\rho_\mathrm{loc}}{\rho_\mathrm{sat}} \right)^{1/\gamma_\mathrm{sp}},
\end{equation}
where $M_\chi$ is the DM particle mass, $\langle \sigma v \rangle$ is the velocity-averaged annihilation cross section, and we assume a benchmark system age $\tau_\mathrm{BH} \approx 10^{10}$ yr. The full DM density profiles for A0620--00 and XTE J1118+480 are shown in Figure \ref{fig:rho_StMBHs}.

A critical factor for the detectability of these mini-spikes is their survivability against tidal disruption within the Galactic environment. As sBHs transit the Galactic disk, gravitational interactions with stars and compact objects tidally strip the outer layers of the DM subhalo. Following the prescription in Ref.~\cite{chanda2024improved}, we estimate a cutoff radius $R_\mathrm{cut}$ beyond which the spike is effectively truncated. Assuming a typical stellar density and velocity dispersion in the disk, the characteristic closest encounter distance $d \sim 10^{-3}$~pc leads to a truncation at
\begin{equation}
    R_\mathrm{cut} \approx d \left( \frac{2M_\mathrm{BH}}{1\,M_\odot} \right)^{1/3}.
\end{equation}
For $r > R_\mathrm{cut}$, the DM density reverts to the ambient local value $\rho_\mathrm{loc} \equiv \rho_{\mathrm{Ein}}(r_{\mathrm{BH}})$. As illustrated in Figure~\ref{fig:rho_StMBHs}, while this truncation significantly reduces the total DM mass in the outer regions of the spike, the densest central regions (which provide the majority of the annihilation signal) remain largely shielded. The resulting parameters for A0620--00 and XTE J1118+480 are detailed in Table~\ref{tab: black hole data}.

\begin{table}[htbp]\label{table:stmbh_data}
    \centering
    \begin{tabular}{|c|c|c|}
        \hline
        \ & \textbf{A0620--00} & \textbf{XTE J1118+480} \\
        \hline
        $M_\mathrm{BH}$~$(M_\odot)$ & $5.86 \pm 0.24$ \cite{van2017mass} & 7.46 $^{+0.34}_{-0.69}$ \cite{gonzalez2013fast} \\
        % \hline
        $D$~(pc) & 1050$\pm$400 \cite{guenette2009veritas} & 1720$\pm$100 \cite{gelino2006inclination}  \\
        % \hline
        $l$~(deg) & 209.96 \cite{liu2007catalogue} & 157.66 \cite{fender2001spectral} \\
        % \hline
        $b$~(deg) & -6.5399 \cite{liu2007catalogue} & 62.321 \cite{fender2001spectral} \\
        \hline
        $R_\mathrm{sp}~(\mathrm{pc})$ & 4.39 & 4.73 \\
        $R_\mathrm{cut}~(\mathrm{pc})$ & $2.27\times 10^{-3}$ & $2.46\times 10^{-3}$ \\
        $\rho_\mathrm{loc}~(\mathrm{GeV/cm}^3)$ & 0.314 & 0.320 \\
        \hline
    \end{tabular}
    \caption{Measured (top) and derived (bottom) parameters of the sBHs A0620--00 and XTE J1118+480.}
    \label{tab: black hole data}
\end{table}

% \begin{table}[htbp]\label{table:stmbh_data}
%     \centering
%     \begin{tabular}{|c|c|c|}
%         \hline
%         \ & \textbf{A0620--00} & \textbf{XTE J1118+480} \\
%         \hline
%         $R_\mathrm{sp}~(\mathrm{pc})$ & 4.39 & 4.73 \\
%         $R_\mathrm{cut}~(\mathrm{pc})$ & $2.27\times 10^{-3}$ & $2.46\times 10^{-3}$ \\
%         $\rho_\mathrm{loc}~(\mathrm{GeV/cm}^3)$ & 0.314 & 0.320 \\
%         \hline
%     \end{tabular}
%     \caption{Measured (top) and derived (bottom) parameters of the sBHs A0620--00 and XTE J1118+480.}
%     \label{tab: black hole data}
% \end{table}

% \begin{table}[htbp]\label{table:stmbh_data}
%     \centering
%     \begin{tabular}{|c|c|c|}
%         \hline
%         \ & \textbf{A0620--00} & \textbf{XTE J1118+480} \\
%         \hline
%         $\dot P$~$(\mathrm{ms\ yr}^{-1})$ & $-0.60 \pm 0.08$ & $-1.90\pm 0.57$   \\
%         \hline
%     \end{tabular}
% \end{table}

\subsection{Gamma-ray Flux and J-factor Determination}
\label{subsection: J-factor}

The expected observable from dark matter annihilation is the differential photon flux, which depends on both the particle physics properties of the WIMP and its astrophysical distribution. It is expressed as
\begin{equation}
    \frac{\mathrm{d}\Phi_\gamma}{\mathrm{d}E} \equiv \frac{\langle \sigma v \rangle}{8\pi M_\chi^2} \, J \sum_i B_i \frac{\mathrm{d} N_i}{\mathrm{d}E_\gamma},
    \label{fluxdiff}
\end{equation}
where $\mathrm{d}N_i/\mathrm{d}E_\gamma$ denotes the photon spectrum for the $i$-th annihilation channel, and $B_i$ is the corresponding branching ratio. The $J$-factor encodes the astrophysical dependence and is defined by the integration of the square of the DM density over the solid angle $\Delta \Omega$ and along the line of sight (l.o.s),

\begin{equation}
    J \equiv \int_{\Delta \Omega} \mathrm{d}\Omega \int_\mathrm{l.o.s} \mathrm{d}l \, \rho^2(r(l, \varphi)).
\end{equation}

Given that the distance to the BHs is much larger than the spike radius ($D \gg R_\mathrm{sp}$), we adopt a point-like approximation. Furthermore, accounting for the tidal disruption discussed in the previous section, the annihilation signal is dominated by the region within the cut-off radius $R_\mathrm{cut}$. The $J$-factor is thus approximated as
\begin{equation}
    J \simeq \frac{4\pi}{D^2} \int_{2R_\mathrm{Sch}}^{R_\mathrm{cut}} \rho^2(r) r^2 \mathrm{d}r.
\end{equation}

This integral can be evaluated analytically, which is particularly advantageous for computational efficiency when scanning the large parameter space of $M_\chi$ and $\langle \sigma v \rangle$. For the case where the cut-off occurs within the spike region ($R_\mathrm{sat} \leq R_\mathrm{cut} \leq R_\mathrm{sp}$), evaluating the integral yields the closed-form expression

\begin{equation}
    J = \frac{4\pi}{D^2} \rho_\mathrm{loc}^{4/3}\rho_\mathrm{sat}^{2/3} R_\mathrm{sp}^{3} 
    \left[ \frac{7}{6} - \frac{2}{3}\left( \frac{R_\mathrm{sat}}{R_\mathrm{cut}} \right)^{3/2} 
    - \frac{1}{2} \left( \frac{2R_\mathrm{Sch}}{R_\mathrm{sat}} \right)^{2} \right].
\end{equation}

Since $\rho_\mathrm{sat}$ depends on the WIMP properties and $R_\mathrm{sp}$ is determined by the BH mass, the $J$-factor is a fully analytical function of the parameters $(D,\ \rho_\mathrm{loc},\ M_\mathrm{BH},\ M_\chi,\ \langle \sigma v \rangle)$. Table~\ref{tab:J-factor} lists the calculated $J$-factors for our targets across three benchmark DM masses, assuming the thermal relic annihilation cross section $\langle \sigma v \rangle = 3 \times 10^{-26} \, \text{cm}^3\text{s}^{-1}$.

The resulting $J$-factors reach $\sim 10^{22} \, \text{GeV}^2/\text{cm}^5$ (see Table~\ref{tab:J-factor}), significantly exceeding those of traditional targets like dwarf spheroidal galaxies (see Refs.~\cite{abramowski2014search,strigari2018dark}). In Table~\ref{tab:J-factor}, we compare $J$-factors calculated with and without the inclusion of the tidal disruption cut-off. We find that the effect of stripping is minor: the reduction in $J$ is approximately 5\% for $M_\chi = 10$~GeV and becomes negligible ($<1\%$) for $M_\chi \geq 100$~GeV. This confirms that the gamma-ray signal is heavily dominated by the densest, central part of the spike, which remains shielded from tidal forces. These $J$-factors serve as the astrophysical input for the likelihood analysis of the \textit{Fermi}-LAT data described in the following section.

\begin{table}[htbp]
    \centering
    \begin{tabular}{|c|c|c|}
        \hline
        $M_\chi$~(GeV) & \textbf{A0620--00} & \textbf{XTE J1118+480} \\
        \hline
        $10$ & $7.72$ & $3.68$ \\
        $100$ & $35.9$ & $17.1$ \\
        $1000$ & $166$ & $79.2$ \\
        \hline
    \end{tabular}
    \caption{Calculated J-factor for the sBHs A0620--00 and XTE J1118+480, for three benchmark DM masses and annihilation cross section fixed to the thermal-relic value. J-factor values are given in units of $10^{20}~\mathrm{GeV^2/cm^5}$.}
    \label{tab:J-factor}
\end{table}

%\begin{figure*}[ht!]
%    \centering
%    \includegraphics[width=\textwidth]{images/J-factor_colormap_2025_11_25.pdf} 
%    \caption{J-factor as function of DM annihilation cross-section (top) and sBH distance (bottom) for DM masses between $1$~GeV and $10$~TeV, for both sBHs in our study, A0620--00 (left) and XTE J1118+480 (right).}
    \label{fig:j-factor_colormaps}
%\end{figure*}

%\begin{figure*}[ht!]
%    \centering
%    \includegraphics[width=\textwidth]{images/cut_J-factor_colormap_2025_11_25.pdf} 
%    \caption{Ratio between J-factor with and without cutoff due to tidal disruption, also as function of DM annihilation cross-section (top) and sBH distance (bottom), for DM masses between $1$~GeV and $10$~TeV, and for both sBHs in our study, A0620--00 (left) and XTE J1118+480 (right).}
%    \label{fig:cut_j-factor_colormaps}
%\end{figure*}

\section{\textit{FERMI}-LAT DATA ANALYSIS}

The \textit{Fermi}-LAT is a gamma-ray pair-conversion detector on board the \textit{Fermi} satellite, in continuous operation since August 2008 \cite{Atwood2009_LAT}. The instrument has a field of view of approximately $2.4$ sr, with an angular resolution that improves significantly with energy, ranging from $\sim 5^\circ$ at 100~MeV to $\sim 0.1^\circ$ above 100~GeV. For this analysis, the two BHs are modeled as point-like sources. We employ the \texttt{P8R3\_SOURCE\_V3} event class, selecting photons in the energy range between 500~MeV and 1~TeV. 
Our study of A0620--00 and XTE J1118+480 utilizes the most recent catalogs provided by the \textit{Fermi}-LAT Collaboration, including the Galactic diffuse emission model (\texttt{gll\_iem\_v07.fits}) and the 4FGL-DR3  point source catalog (\texttt{gll\_psc\_v35.fits})\cite{Fermi-LAT:2022byn}\cite{Fermi-LAT:2019yla}. 

The analysis was performed using \texttt{fermipy}(v1.2) \cite{2017ICRC...35..824W}, which is based on the underlying \texttt{Fermitools} package (v2.2.0)~\footnote{\url{https://fermi.gsfc.nasa.gov/ssc/data/analysis/software/}}. We employed eight energy bins per decade, defined on a logarithmic scale, and a spatial pixel size of $0.08^\circ$. To mitigate contamination from photons produced in the atmosphere of the Earth, a maximum cut in the zenith angle of $100^\circ$ was applied. For each target position, a region of interest (ROI) of $10^\circ \times 10^\circ$ was defined. Taking into account the broadening of the detector’s PSF and the resulting leakage of photons from sources outside the ROI, the model included all sources listed in the 4FGL-DR3 catalog  within an extended region of $15^\circ \times 15^\circ$ \cite{mcdaniel2023legacyanalysisdarkmatter}.
The standard statistical analysis in gamma-ray astronomy is based on the \textit{test statistic} (TS), defined as $TS=2\ln(\mathcal{L}_1/\mathcal{L}_0)$ where $\mathcal{L}_0$ is the likelihood function of the null hypothesis (background-only) and $\mathcal{L}_1$ that of the alternative hypothesis (background plus a point source) \cite{Wilks1938}. \texttt{fermipy} computes the TS by incorporating the Galactic diffuse emission, the isotropic component, and all identified 4FGL sources. This allows for the generation of \textit{residual TS maps} around A0620--00 and XTE J1118+480 to identify potential excesses.

\begin{figure}[b!]
    \centering
    \includegraphics[width=0.4\textwidth]{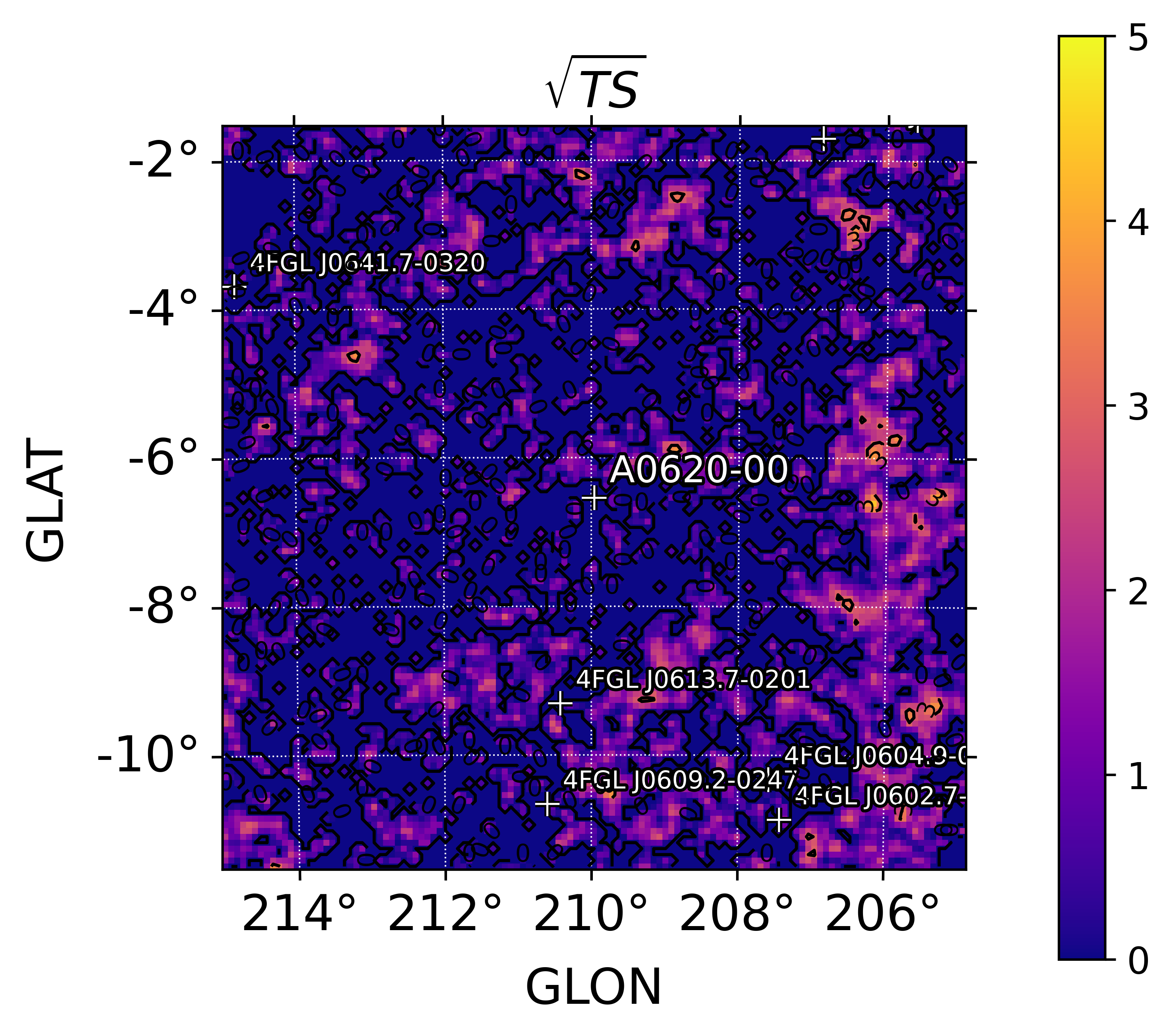}
    \caption{Residual TS map centered on the black hole A0620--00, covering an area of $10^\circ \times 10^\circ$. We find no statistically significant excess in 17 years of \textit{Fermi}-LAT observations, neither in the point-source nor in the extended-source analysis at the position of A0620--00.}
    \label{mapaTS1}
\end{figure}
\begin{figure}[t!]
    \centering
    \includegraphics[width=0.4\textwidth]{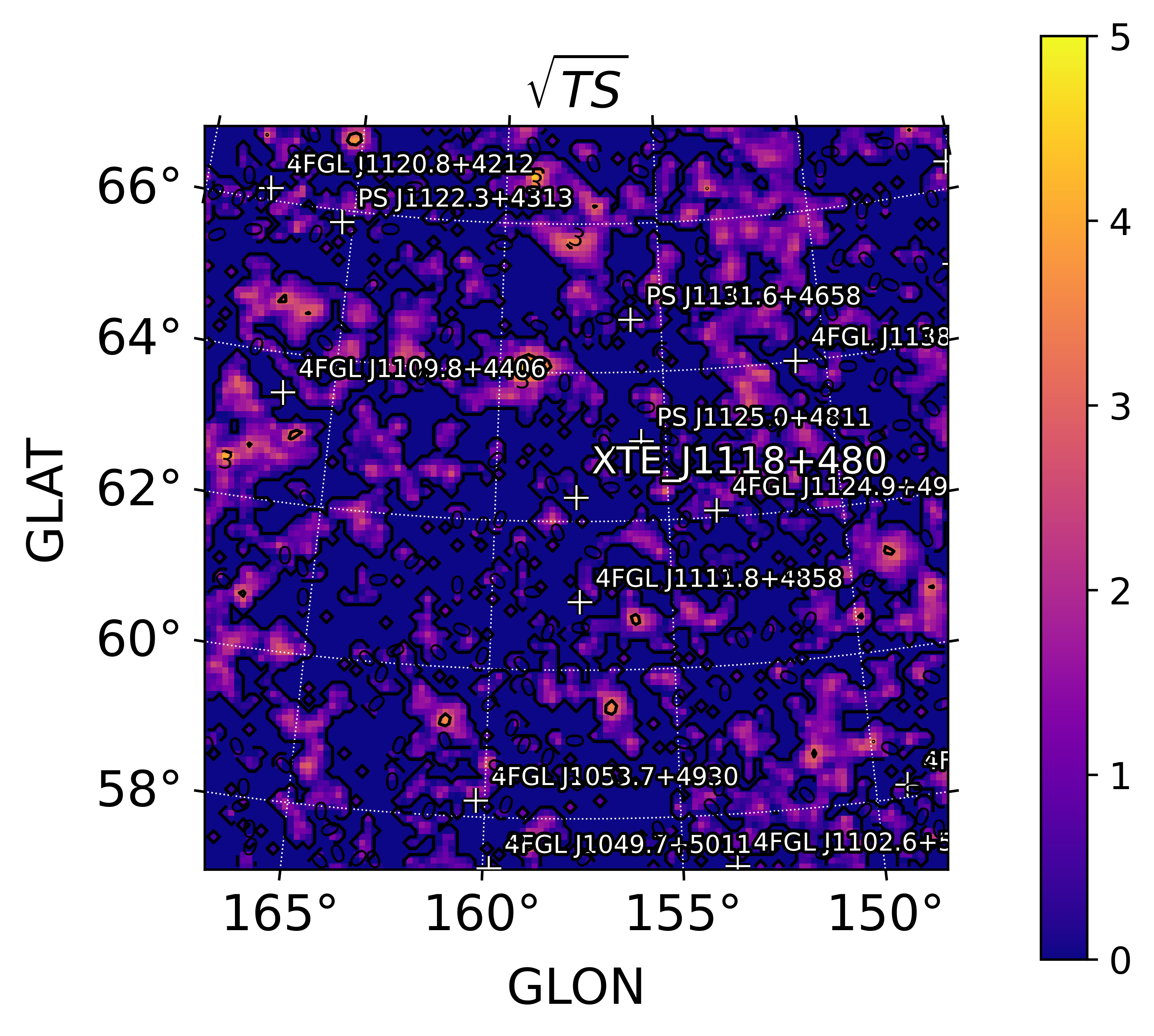}
    \caption{Residual TS map centered on the black hole XTE J1118+480, covering an area of $10^\circ \times 10^\circ$. We find no statistically significant excess in 17 years of \textit{Fermi}-LAT observations, neither in the point-source nor in the extended-source analysis at the position of XTE J1118+480.}
    \label{mapaTS2}
\end{figure}

Spectral parameters of sources within $5^\circ$ of the ROI center were left free to vary, while those beyond this radius were fixed to their 4FGL-DR3 catalog values. The normalization of both the Galactic and isotropic diffuse components were also left free to vary during the likelihood minimization.

Using the \texttt{GTAnalysis} tool from \texttt{fermipy}, we computed the spectral energy distribution (SED) of both black holes using the \texttt{sed()} method. This procedure conducts a maximum likelihood fit in each independent energy bin assuming a power-law distribution with a fixed spectral index of $\Gamma = 2$. The output of the \texttt{sed()} method provides the likelihood profile as a function of the differential gamma-ray flux in each bin, $\Delta \ln \mathcal{L}_i (\mathrm{d}\Phi_\gamma/\mathrm{d}E, E_i)$.
\begin{figure}[b!]
    \centering
\includegraphics[width=0.4\textwidth]{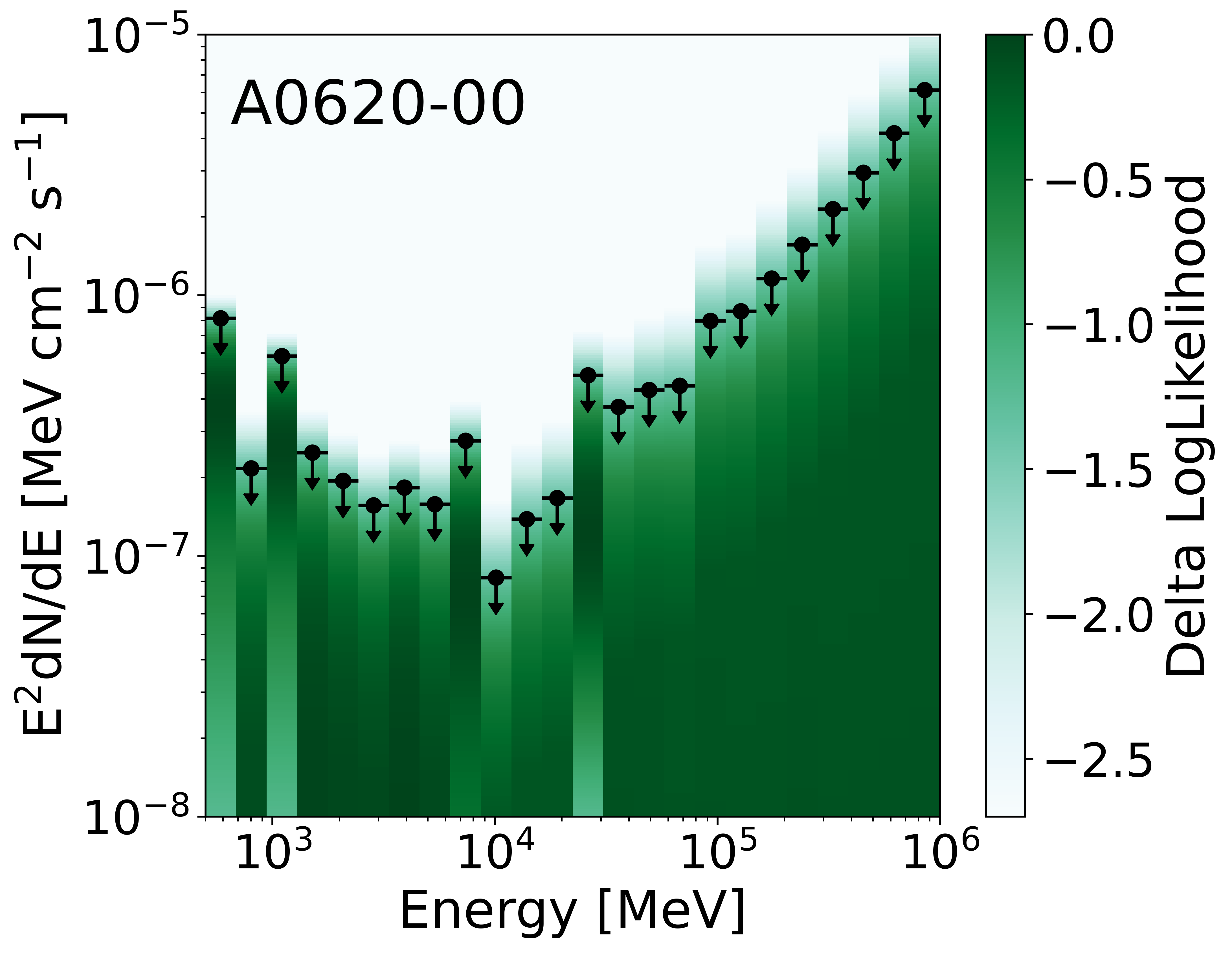}
    \caption{95\% CL upper limits on the flux for the source A0620--00. The spectral energy distribution and the likelihood profile are obtained in the energy range from 500~MeV to 1~TeV, assuming a point-like source morphology.}
    \label{fig:loglike_A0620-00}
\end{figure}

\begin{figure}[b!]
    \centering
\includegraphics[width=0.4\textwidth]{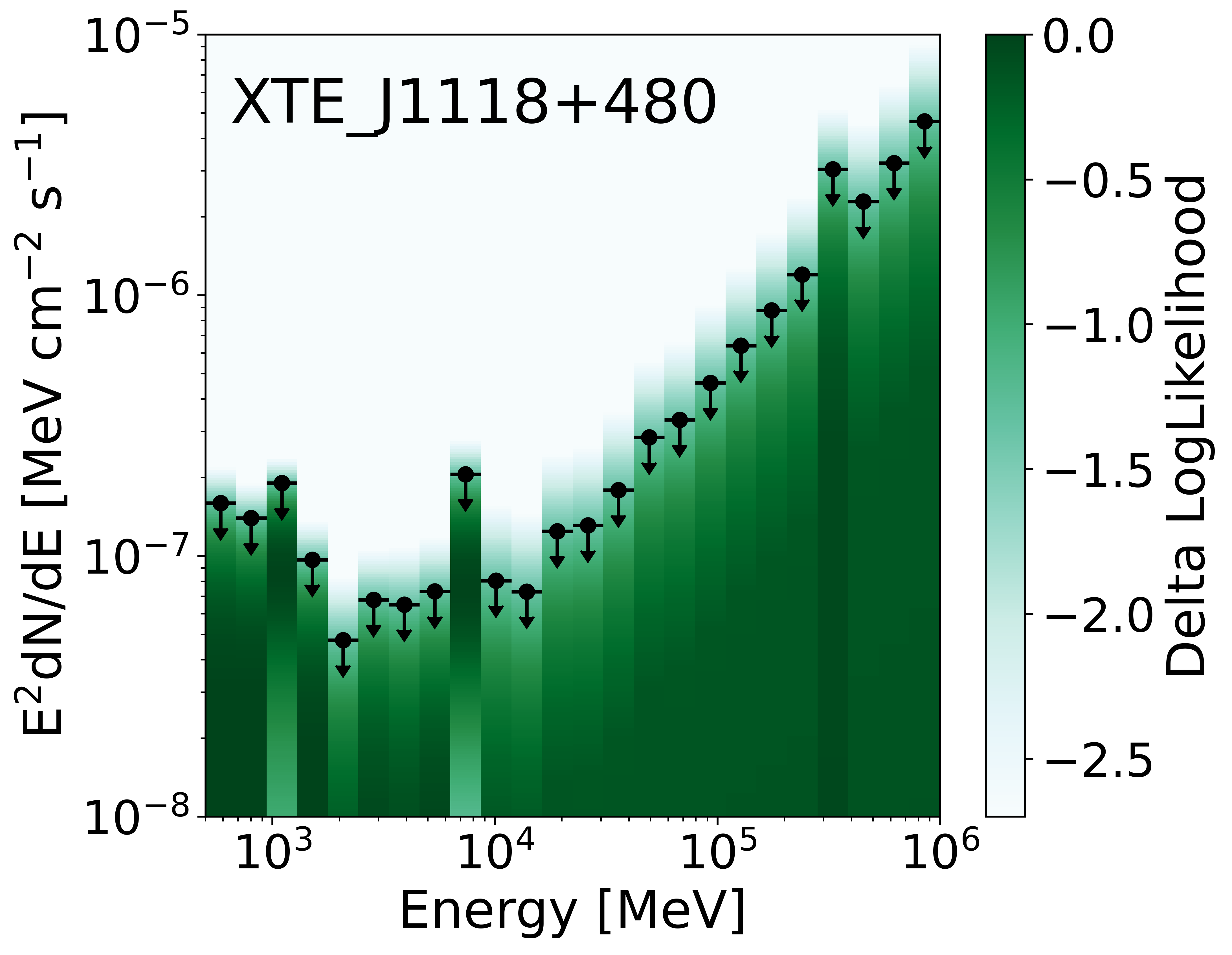}
    \caption{95\% confidence level upper limits on the flux are shown for the source XTE J1118+480. The spectral energy distribution and the likelihood profile are obtained in the energy range from 500~MeV to 1~TeV, assuming a point-like source morphology.}
    \label{fig:loglike_XTE_J1118+480}
\end{figure}

This framework allows us to recast the observational data into the parameter space. The differential gamma-ray flux expected from DM annihilation is given by Eq.~(\ref{fluxdiff}). To constrain the DM properties, we compute the joint TS as a function of particle mass $M_\chi$ and annihilation cross section $\langle \sigma v \rangle$:

\begin{equation}
\begin{split}
TS(M_\chi, \langle \sigma v \rangle) ={}&
-2 \sum_i \Bigg(
\ln \mathcal{L}_0 \\
&\quad - \ln \mathcal{L}_i \left[
\frac{\mathrm{d}\Phi_\chi}{\mathrm{d}E}
(M_\chi,\langle \sigma v \rangle,E_i)
\right]
\Bigg)
\end{split}
\end{equation}
% \begin{equation}
%     TS(M_\chi, \langle \sigma v \rangle) = -2 \sum_{i} \left( \ln \mathcal{L}_0 - \ln \mathcal{L}_i \left[ \frac{\mathrm{d}\Phi_\chi}{\mathrm{d}E} (M_\chi, \langle \sigma v \rangle, E_i) \right] \right),
% \end{equation}
where the sum runs over all energy bins $E_i$. The theoretical differential flux $\mathrm{d}\Phi_\chi/\mathrm{d}E$ is calculated using the DM energy spectra provided by the PPPC4DMID cookbook~\cite{cirelli2011pppc}.

We consider a grid of DM masses between 1~GeV and 10~TeV and scan the cross section values to determine the 95\% confidence level (C.L.) upper limits. For each mass, the limit is defined as the value of $\langle \sigma v \rangle$ for which the TS decreases by 2.71 relative to the maximum likelihood. This analysis is performed for three representative annihilation channels: $b\bar{b}$, $\tau^+\tau^-$, and $W^+W^-$.

Finally, to investigate the broader discovery potential of \textit{Fermi}-LAT for sBH mini-spikes, we derive lower limits on the source distance using the \texttt{fermipy-flux-sensitivity} module. For a fixed annihilation cross section (taken here as the canonical thermal-relic value) the instrumental flux sensitivity maps directly to a distance boundary. This mapping is governed by the $J$-factor dependence on the local DM density,  $\rho_{\rm loc}$, and its inverse-square scaling with the source distance, $1/D^2$. The resulting distance thresholds exhibit significant spatial variation, primarily driven by the intensity of the astrophysical background (most notably the Galactic diffuse emission) which dictates the local signal-to-noise ratio. To a lesser extent, these limits also incorporate the energy-dependent effective area and Point Spread Function (PSF) of the LAT, which determine the instrument's ability to resolve the DM annihilation spectrum against the local background at each sky position.

\section{Results}

The standard statistical analysis of 17 years of \textit{Fermi}-LAT data reveals no significant gamma-ray emission from the stellar-mass black holes A0620--00 and XTE J1118+480. The residual TS maps, centered on the target coordinates and covering a $10^\circ \times 10^\circ$ area, are shown in Figures~\ref{mapaTS1} and~\ref{mapaTS2}. In both cases, the maximum test statistic at the source position is $TS < 1$, a result fully consistent with the background-only null hypothesis.

In the absence of a detectable signal, we first calculate the 95\% C.L. upper limits on the differential gamma-ray flux at the positions of the two sources, assuming point-source morphology. The resulting spectral energy distributions and likelihood profiles as a function of differential flux and energy are presented in Figures~\ref{fig:loglike_A0620-00} and~\ref{fig:loglike_XTE_J1118+480}.

\begin{figure}[htbp]
    \centering
    \includegraphics[width=0.5\textwidth]{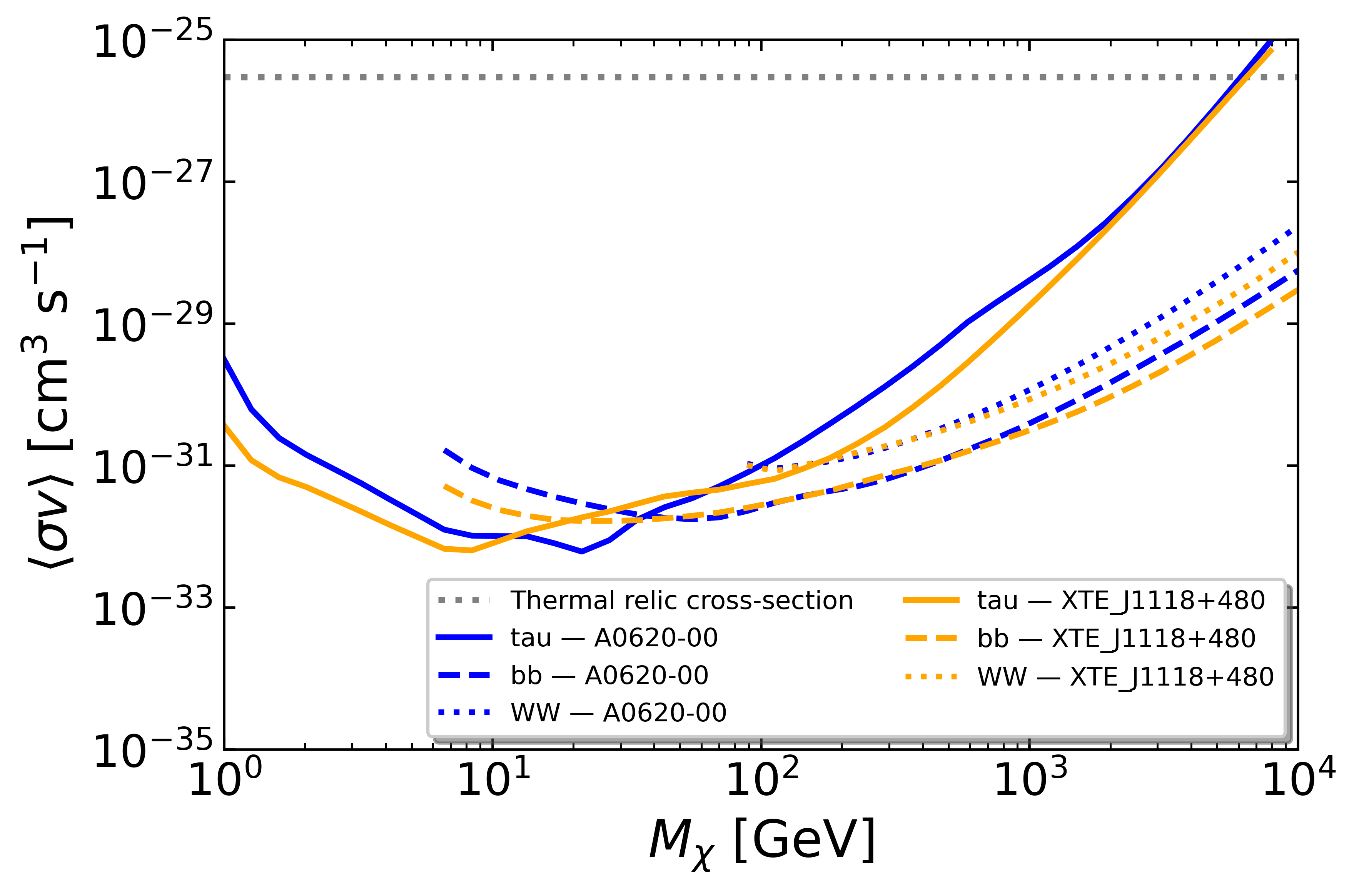}
    \caption{95\% C.L. upper limits on the DM annihilation cross section ($\langle \sigma v \rangle$, in cm$^3$s$^{-1}$) as function of the DM particle mass for the annihilation into $b\bar{b}$ (solid line), $\tau^{+}\tau^{-}$  (dashed line) and  $W^{+}W^{-}$ (dotted line) channels using 17 years of observation by \textit{Fermi}-LAT of the sources A0620--00 (blue curve) and XTE~J1118+480 (orange curve). The limits were derived assuming a 100\% branching ratio for each of the considered annihilation channels. The grey dashed line represents the thermal relic cross section.}
    \label{fig:cross_section_upper_limits}
\end{figure}

\begin{table}[htbp]
\centering
\caption{95\% C.L. upper limits on the DM annihilation cross section ($\langle \sigma v \rangle$, in cm$^3$s$^{-1}$).}
\label{tab:dm_limits}
% O resizebox garante que a tabela caiba na coluna do JCAP
\resizebox{\columnwidth}{!}{%
\begin{tabular}{l|c|c|c} % As barras "|" criam as linhas verticais
\hline
\textbf{Channel} & \textbf{100 GeV} & \textbf{1 TeV} & \textbf{10 TeV} \\
\hline % Linha única separando o cabeçalho
\multicolumn{4}{c}{\textbf{A0620--00}} \\ 
\hline

$W^+W^-$       & $9.28 \times 10^{-32}$ & $1.1 \times 10^{-30}$ & $2.23 \times 10^{-28}$ \\
$b\bar{b}$     & $2.48 \times 10^{-32}$ & $3.71 \times 10^{-31}$ & $5.34 \times 10^{-29}$ \\
$\tau^+\tau^-$ & $9.52 \times 10^{-32}$ & $3.87 \times 10^{-29}$ & $2.69 \times 10^{-25}$ \\
\hline
\multicolumn{4}{c}{\textbf{XTE~J1118+480}} \\ 
\hline
$W^+W^-$       & $8.73 \times 10^{-32}$ & $8.46 \times 10^{-31}$ & $9.64 \times 10^{-29}$ \\
$b\bar{b}$     & $2.63 \times 10^{-32}$ & $3.11 \times 10^{-31}$ & $2.88 \times 10^{-29}$ \\
$\tau^+\tau^-$ & $6.36 \times 10^{-32}$ & $1.75 \times 10^{-29}$ & $1.75 \times 10^{-25}$ \\
\hline
\end{tabular}%
}
\end{table}

Utilizing the analytical $J$-factors derived in Section~\ref{subsection: J-factor}, we translate the flux likelihoods into constraints on the DM parameter space. Figure~\ref{fig:cross_section_upper_limits} presents the 95\% C.L. upper limits on the velocity-averaged DM annihilation cross section, $\langle \sigma v \rangle$, as a function of the particle mass $M_\chi$. We consider 100\% branching ratios for the $b\bar{b}$ (solid line), $\tau^{+}\tau^{-}$ (dashed line), and $W^{+}W^{-}$ (dotted line) annihilation channels for both A0620--00 (blue curves) and XTE~J1118+480 (orange curves). 

The results demonstrate that the thermal relic WIMP cross section ($3 \times 10^{-26} \, \text{cm}^3\text{s}^{-1}$) is excluded across the entire mass range considered (10~GeV to 10~TeV) for the $b\bar{b}$ and $W^{+}W^{-}$ channels. For the $\tau^{+}\tau^{-}$ channel, the exclusion power remains significant, ruling out the thermal relic cross section for all masses below $\sim$6~TeV. The numerical upper limits for representative benchmark masses are provided in TABLE~\ref{tab:dm_limits}.

\begin{figure*}[htbp]
    \centering
    \includegraphics[width=0.8 \textwidth]{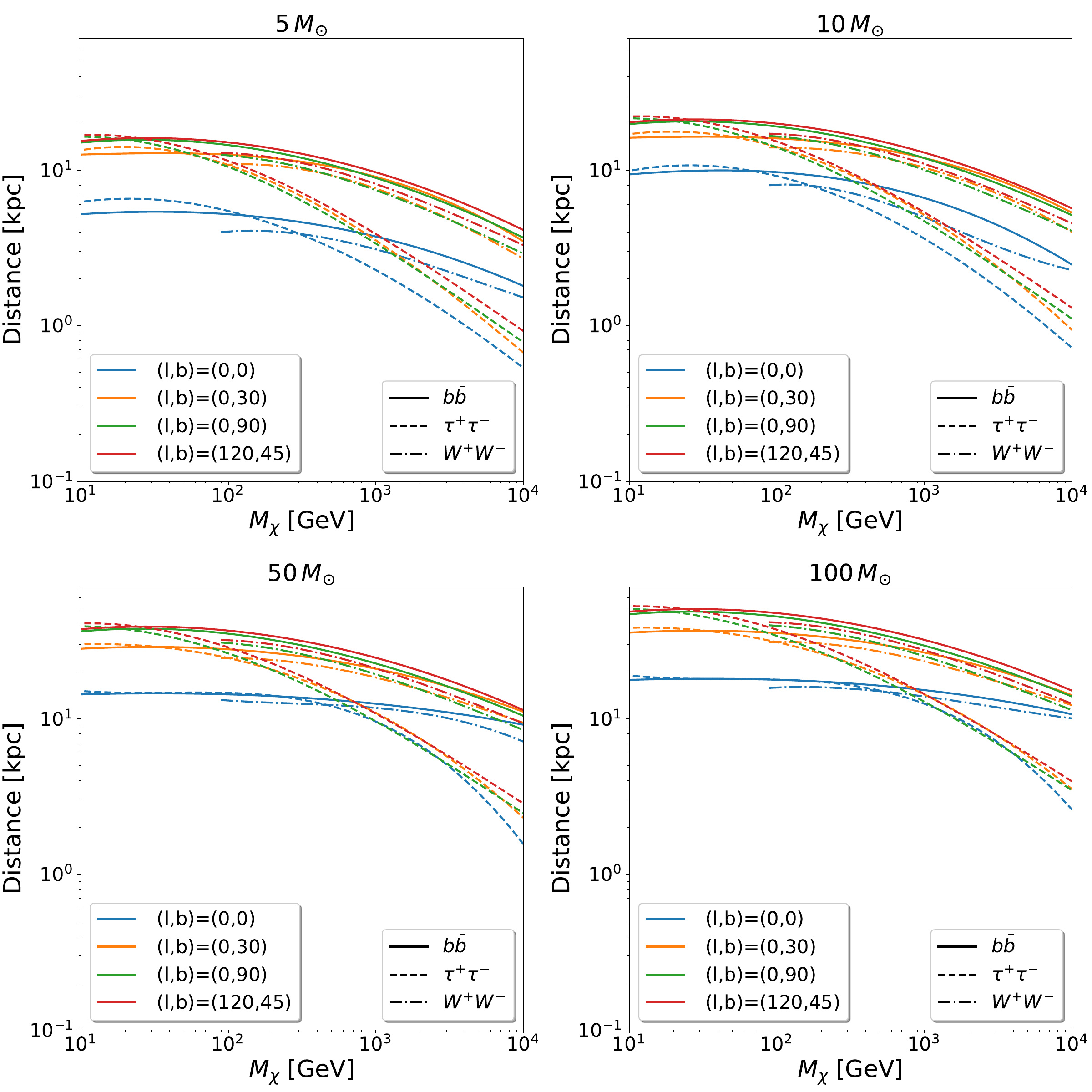}
    \caption{Lower limits on the source distance are presented as a function of the DM particle mass, $M_\chi$, corresponding to a 95\% C.L. These constraints are derived from the \textit{Fermi}-LAT flux sensitivity assuming a point-like source and a thermal relic cross-section of $\langle\sigma v\rangle = 3\times10^{-26}\,\mathrm{cm^3\,s^{-1}}$. The four panels correspond to scenarios with BH masses of $5\,M_\odot$, $10\,M_\odot$, $50\,M_\odot$, and $100\,M_\odot$. The color coding indicates four different Galactic coordinates to illustrate the effect of the background level. Finally, the results were obtained for distinct annihilation channels, $b\bar{b}$ (solid curves), $\tau^{+}\tau^{-}$ (dashed curves), and $W^{+}W^{-}$ (dotted curves).}
    \label{fig:distance_lower_limits}
\end{figure*}

Finally, to investigate the broader discovery potential of \textit{Fermi}-LAT for sBH mini-spikes, we estimate the detectability reach for a set of generic stellar-mass black holes with masses of $5\,M_\odot$, $10\,M_\odot$, $50\,M_\odot$, and $100\,M_\odot$. In Figure~\ref{fig:distance_lower_limits}, we present the 95\% C.L. lower limits on the source distance as a function of $M_\chi$, assuming the thermal relic annihilation cross section. These limits were derived for four different sky locations to illustrate the impact of varying background levels: the Galactic Center $(0^\circ,0^\circ)$, a region off the Galactic Plane $(0^\circ,30^\circ)$, the North Galactic Pole $(0^\circ,90^\circ)$, and an extragalactic field at $(120^\circ,45^\circ)$. 

The derived distance limits exhibit a strong dependence on both the astrophysical environment and the physical scale of the system. As shown in Figure~\ref{fig:distance_lower_limits}, the discovery reach in low-background extragalactic fields, such as the North Galactic Pole and the $(120^\circ,45^\circ)$ direction, can extend beyond 20~kpc for sBHs with $M_{\text{BH}} \geq 10\,M_\odot$. Notably, even for masses as low as $5\,M_\odot$, the \textit{Fermi}-LAT maintains the sensitivity to detect a WIMP annihilation spike toward the Galactic Center at distances as far as $\sim 3$~kpc (for $b\bar{b}$ and $W^+W^-$ channels) or $\sim 2$~kpc (for $\tau^+\tau^-$). In extragalactic fields, this reach for a $5\,M_\odot$ sBH increases to approximately $8$~kpc for the $b\bar{b}$ and $W^+W^-$ channels, and $\sim 3$~kpc for the $\tau^+\tau^-$ channel. Furthermore, for sBHs more massive than a few $10\,M_\odot$, the \textit{Fermi}-LAT remains sensitive to mini-spikes for WIMP masses below 1~TeV even at distances surpassing that of the Galactic Center itself, demonstrating that the search is sensitive to a significant fraction of the Galactic sBH population.
\section{Discussion and Conclusions}

In this work, we have performed a targeted search for gamma-ray signatures from dark matter mini-spikes around the stellar-mass black holes A0620--00 and XTE J1118+480 using 17 years of \textit{Fermi}-LAT observations. This search was motivated by the proposal that the anomalous orbital decay measured in these systems is driven by dynamical friction within dense dark-matter enhancements. By adopting a framework of adiabatic spike formation, while accounting for the dynamical reality of tidal stripping in the Galactic disk, we modeled each target as a point-like source and quantified the expected signal via derived $J$-factors.

Our null results provide a robust observational test of the hypothesis that these sBHs are primordial and embedded in a thermal WIMP halo. If the observed orbital decay in A0620--00 and XTE J1118+480 were indeed caused by the presence of dense DM spikes, the resulting gamma-ray luminosity would have reached the \textit{Fermi}-LAT detection threshold across the majority of the Galactic volume. The absence of such a signal creates a fundamental tension, suggesting that either the dark matter does not consist of thermal-relic WIMPs, or the dynamical interpretation of the orbital decay must be reconsidered.

These findings reinforce the evidence for a significant incompatibility between stellar-mass primordial BHs and the WIMP paradigm~\cite{adamek2019wimps,Lacki:2010zf,Boucenna:2017ghj}. If PBHs exist in this mass range, they must inhabit a Universe where the DM candidate either does not annihilate, or possesses a cross-section significantly below the thermal-relic value. Furthermore, our recast of the distance reach demonstrates the impressive sensitivity of the \textit{Fermi}-LAT to this paradigm. Even for sBHs with masses as low as $5\,M_\odot$, the instrument is capable of detecting mini-spikes at distances of $\sim 3$~kpc toward the Galactic Center, and as far as $\sim 8$~kpc in extragalactic fields. For sBHs more massive than $10\,M_\odot$, the reach extends beyond the Galactic Center itself, depending on the WIMP mass and annihilation channel.

Finally, the potential of this approach will expand as the observational sample of sBHs grows. The identification and characterization of additional candidates will allow the application of this same analysis pipeline to a larger ensemble of systems. Such a population-based study will enhance overall sensitivity, mitigate source-specific backgrounds, and further strengthen the utility of gamma-ray observations as a definitive probe of dark matter behavior in the extreme environments surrounding black holes.

\section*{Acknowledgements} 
Authors are supported by the S\~{a}o Paulo Research Foundation (FAPESP) through grants number 2019/14893-3, 2020/00320-9, 2021/01089-1. AV is supported by FAPESP grant No 2024/15560-6, and CNPq grant 309613/2025-6. AVAMB is supported by the Brazilian Federal Agency for Support and Evaluation of Graduate Education (CAPES) through grant No 88887.803875/2023-00. MGGS is supported by CAPES through grant No 88887.939389/2024-00 and by FAPESP through grant No 2024/01381-2. The authors acknowledge the National Laboratory for Scientific Computing (LNCC/MCTI,  Brazil) for providing HPC resources for the SDumont supercomputer (http://sdumont.lncc.br).

\bibliography{references}

\end{document}